# An algorithm for dividing two complex numbers

Aleksandr Cariow

*West Pomeranian University of Technology, Szczecin, Faculty of Computer Science and Information Technology, Żołnierska 49, Szczecin 71-210, Poland*

**Abstract:**

*In this work a rationalized algorithm for calculating the quotient of two complex numbers is presented which reduces the number of underlying real multiplications. The performing of a complex number division using the naive method takes 4 multiplications, 3 additions, 2 squarings and 2 divisions of real numbers while the proposed algorithm can compute the same result in only 3 multiplications (or multipliers – in hardware implementation case), 6 additions, 2 squarings and 2 divisions of real numbers.*

**Keywords:**

*Complex number division, Gauss-like algorithmic trick, reduction of number of real multiplications*

## 1. Introduction

Mathematical calculations with complex numbers are widely used today for data processing in numerous fields of science and engineering. Perhaps it is easier to list the fields where these numbers are not applicable than those where they are applied.

In complex-valued arithmetic the most time and area consuming operations are multiplication and division of two complex numbers. What is more, the division is even more complicated and expensive than multiplication [1-3].

The schoolbook multiplication of complex numbers requires performing 4 real multiplications and 2 real additions, and the schoolbook division of complex numbers requires performing 4 real multiplications, 2 real squarings, 3 real additions and 2 real divisions. In turn, multiplication and division of two ordinary real numbers are also more time consuming operations than addition or subtraction of ordinary real numbers. Since these operations are carried out repeatedly, the total time to the correct implementation of the final algorithm may be unacceptable. It is therefore evident that finding ways, which reduce the number of real multiplications required for performing multiplication and division of complex numbers, is very important task. In 1805 Gauss had discovered a way (so called Gauss's complex multiplication trick) of reducing the number of multiplications to three [4, 5]. The Gauss's optimization saves one real multiplication out of four. Unfortunately, no such solutions for division of the complex numbers have been proposed. The aim of the present paper is to suggest an efficient algorithm for this purpose.

## 2. The algorithm

Let us suppose that, we need to divide two complex numbers

$$y = \frac{a}{x} \qquad (1)$$

where $a = a_r + ia_i$, $x = x_r + ix_i$, and $y = y_r + iy_i$ are complex numbers; $x_r$, $x_i$, $y_r$, $y_i$ are real variables, and $i$ is the imaginary unit, satisfying $i^2 = -1$. Subscript $r$ means the real part of complex number, and the subscript $i$ means the imaginary part of complex number.

The task is to calculate the quotient defined by the expression (1) with the minimal multiplicative complexity.

A schoolbook method of finding the quotient of two complex numbers can be represented by the following equations:

$$y_r = \frac{a_r x_r + a_i x_i}{x_r^2 + x_i^2}, \quad y_i = \frac{a_i x_r - a_r x_i}{x_r^2 + x_i^2} \qquad (2)$$

Direct implementation of the calculations in accordance with these equations requires 4 multiplications, 3 additions, 2 squarings and 2 divisions of real numbers. Below we will show how you can reduce the multiplicative complexity of the operation of calculating the quotient of two complex numbers.

Let us introduce the two column vectors: $\mathbf{X}_{2\times 1} = [x_r, x_i]^T$ and $\mathbf{Y}_{2\times 1} = [y_r, y_i]^T$.

Let also $R = x_r^2 + x_i^2$,

Now the division of complex numbers can be presented in the matrix–vector multiplication form as

$$\mathbf{Y}_{2\times 1} = \frac{1}{R} \mathbf{A}_2 \mathbf{X}_{2\times 1} \qquad (3)$$

where

$$\mathbf{A}_2 = \begin{bmatrix} a_r & a_i \\ a_i & -a_r \end{bmatrix}$$

The direct multiplication of the vector–matrix product in Eq. (3) requires 4 real multiplications and 2 real additions. We propose a simple way to reduce multiplicative complexity of this operation to 3 real multiplications at the price of 3 more real additions. Proposed way will be quite similar to the Gauss's trick for complex number multiplication.

It is easily verify [6] that the matrix with this structure can be factorized, than the computational procedure for calculate the product of $\mathbf{A}_2 \mathbf{X}_{2\times 1}$ can be represented as follows:

$$\mathbf{Y}_{2\times 1} = \mathbf{D}_2 \mathbf{T}_{2\times 3} \mathbf{D}_3 \mathbf{T}_{3\times 2} \mathbf{X}_{2\times 1}$$

where

$$\mathbf{D}_3 = diag\{(a_r - a_i), -(a_r + a_i), a_i\},$$

$$\mathbf{D}_2 = \begin{bmatrix} \delta & 0 \\ 0 & \delta \end{bmatrix} \mathbf{T}_{2\times 3} = \begin{bmatrix} 1 & 0 & 1 \\ 0 & 1 & 1 \end{bmatrix}, \mathbf{T}_{3\times 2} = \begin{bmatrix} 1 & 0 \\ 0 & -1 \\ 1 & 1 \end{bmatrix}, \text{ and } \frac{1}{R} = \delta$$

Fig. 1 shows a data flow diagram of the rationalized algorithm for computation of the quotient of two complex numbers. Data flow diagram is oriented from left to right. Straight lines in the figure denote the operations of data transfer. Points where lines converge denote summation. The dashed line indicates the sign change operation. We deliberately use the usual lines without arrows on purpose, so as not to clutter the picture. The circles in these figures show the operation of multiplication by a real number inscribed inside a circle.

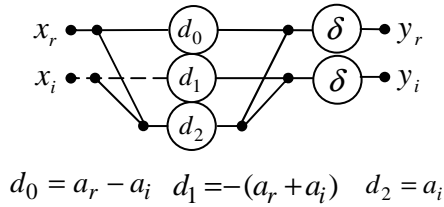

$$d_0 = a_r - a_i \quad d_1 = -(a_r + a_i) \quad d_2 = a_i$$

*Fig.1. Data flow diagram of the rationalized algorithm for computation of the quotient of two complex numbers*

## 3. Conclusion

We presented a hardware oriented algorithm for calculating the quotient of two complex numbers. The performing of a complex number division using the naive method takes 4 multiplications, 3 additions, 2 squarings and 2 divisions of real numbers while the proposed algorithm can compute the same result in only 3 multiplications, 6 additions, 2 squarings and 2 divisions of real numbers. So, the use of this algorithm reduces the multiplication complexity of complex number dividing, thus reducing hardware complexity and leading to a high-speed structure suitable for VLSI implementation. In low power VLSI design, optimization must be primarily done at the level of logic gates amount. From this point of view a multiplication requires much more intensive hardware resources than an addition. Moreover, a multiplier occupies much more area and consumes much more power than an adder. This is because the hardware complexity of a multiplier grows quadratically with operand size, while the hardware complexity of an adder increases linearly with operand size. Therefore, the algorithm for division of complex numbers containing as little as possible of real multiplications is preferable.